\documentclass{llncs}
\usepackage{graphicx}
\usepackage{amsmath}
\usepackage{csquotes}
\usepackage{multirow}
\begin{document}
\title{Unsupervised Intrusion Detection System for Unmanned Aerial Vehicle with Less Labeling Effort}
%\title{Unsupervised Anomaly Detection for Unmanned Aerial Vehicle}

%
%\titlerunning{Abbreviated paper title}
% If the paper title is too long for the running head, you can set
% an abbreviated paper title here
%author

\author{
Kyung Ho Park\and
Eunji Park\and
Huy Kang Kim
}

%\author{
%No author given
%}
%\authorrunning{K. Park et al.}
% First names are abbreviated in the running head.
% If there are more than two authors, 'et al.' is used.
%

\institute{
School of Cybersecurity, Korea University\\%, Republic of Korea\\
\email{\{kyungho96,epark911,cenda\}@korea.ac.kr}}

%\institute{No institute given}

\maketitle  % typeset the header of the contribution

\begin{abstract}
Along with the importance of safety, an IDS has become a significant task in the real world. Prior studies proposed various intrusion detection models for the UAV. Past rule-based approaches provided a concrete baseline IDS model, and the machine learning-based method achieved a precise intrusion detection performance on the UAV with supervised learning models. However, previous methods have room for improvement to be implemented in the real world. Prior methods required a large labeling effort on the dataset, and the model could not identify attacks that were not trained before. 

To jump over these hurdles, we propose an IDS with unsupervised learning. As unsupervised learning does not require labeling, our model let the practitioner not to label every type of attack from the flight data. Moreover, the model can identify an abnormal status of the UAV regardless of the type of attack. We trained an autoencoder with the benign flight data only and checked the model provides a different reconstruction loss at the benign flight and the flight under attack. We discovered that the model produces much higher reconstruction loss with the flight under attack than the benign flight; thus, this reconstruction loss can be utilized to recognize an intrusion to the UAV. With consideration of the computation overhead and the detection performance in the wild, we expect our model can be a concrete and practical baseline IDS on the UAV.

\end{abstract}

%%%%%%%%%%%%%%%%% INTRODUCTION
\section{Introduction}
The Unmanned Aerial Vehicle (UAV) is a promising future technology due to its various applications. The UAVs can deliver packages or medicines at the urgent medical circumstances, or ship goods and products rapidly in the urban areas \cite{pajares2015overview}. Although the UAVs provide a wide range of benefits to the society, however, concerns on the safety and security still exist \cite{choudhary2018intrusion}. If the UAV's communication signals are intruded, UAVs without appropriate control might cause a severe problem. UAVs under attack might not be able to come back to their base station, or they could fail to land on the safe zone under the emergency. Along with the safety concerns, it has become a significant task to identify whether the UAV is intruded or not; thus, an intrusion detection system (IDS) has emerged into the academia and the industry.

An IDS is a security technology that recognizes an intrusion into the computer system \cite{biermann2001comparison}. An IDS on the UAV especially recognizes abnormal patterns or unauthorized activities at the UAV by analyzing activity logs \cite{choudhary2018intrusion}. As the IDS can identify abnormal UAV activities during the flight, several studies it has been researched from the past. The first approach of the IDS on the UAV was a rule-based model. Prior studies analyzed the pattern of UAVs during the flight under attack, and extracted features which describe an abnormal status. The proposed rule-based models achieved a concrete baseline of the IDS on the UAV; however, the model performance was not precise enough to be deployed in a real world. If an abnormal pattern of the UAV is not identified by the detection rules, the proposed models could not recognize the status as an intrusion. Thus, it has become an essential task to increase the detection performance at various types of attacks.

To improve the limit of prior studies, a detection model with machine learning models have been proposed. Numerous machine learning models precisely learn the pattern of UAVs during the flight; thus, presented machine learning-based models achieved a significant detection performance. Although proposed models improved the detection performance from the past, there existed a labeling problem. As prior approaches leveraged supervised machine learning models, a practitioner must provide a well-labeled training data into the model. Under supervised learning, the practitioner should collect and label the flight data under attack, and this data collection process accompanies an enormous cost and effort. Furthermore, a detection model with supervised learning cannot identify non-trained patterns of attack. If a malicious intruder performs non-trained attacks into the UAV, the IDS cannot identify the attack; thus, we analyzed the IDS on the UAV should be implemented without supervised learning.

In this study, we propose a novel IDS for UAVs leveraging unsupervised learning. We presented a series of analyses to extract features from the raw log data and how we transformed it into an effective form. We designed the detection model with an autoencoder, a deep neural network of unsupervised learning, and trained the model with the benign flight data only. Lastly, we validated the model precisely recognizes two types of attack (DoS attack, GPS Spoofing attack) from the benign status. Throughout the study, key contributions are described below:
\begin{itemize}
    \item We designed an intrusion detection model leveraging unsupervised learning with the benign flight data only; thus, our approach reduces a labeling effort.
    
    \item The proposed intrusion detection model effectively recognized the difference between the benign flight and the flight under two types of attack: DoS attack and GPS Spoofing attack.
    
    \item Our study illustrated a series of analysis to produce essential features from raw log data, and extracted features can be applied into the common UAVs.
\end{itemize}

%%%%%%%%%%%%%%%%% LITERATURE REVIEW
\section{Literature Review}
Researchers have proposed various IDS approaches on the UAV. Prior works can be categorized into two streams as shown in Table \ref{papers}: a rule-based approach and the machine learning-based approach.

\begin{table*}[ht]
\centering
\caption{Prior researches of the IDS on UAVs}\label{tab1}
\resizebox{0.95\textwidth}{!}{
\begin{tabular}{|c|c|c|c|}

\hline
\textbf{Category} & \textbf{Intrusion Type} & \textbf{Key Model} & \textbf{Reference} \\
\hline
& SYN Flooding, Password Guessing, & Behavioral Rules & \cite{mitchell2013adaptive}\\
& Buffer Overflow, Scanning & & \\
\cline{2-4}

Rule-based & Spoofing, Gray/Blackhole Attacks & Hierarchical Scheme & \cite{sedjelmaci2017hierarchical}\\
Approach & False Information Dissemination, Jamming & & \\
\cline{2-4}

& Constant Flash-Crowd Attack & Spectral Analysis & \cite{zhang2018network}\\

& Progressive Flash-Crowd Attack & & \\

\hline
& SYN Flooding, Password Guessing, & PSO-DBN & \cite{tan2019intrusion}\\
Machine Learning & Buffer Overflow, Scanning & & \\
\cline{2-4}

Approach & Spoofing, Jamming & STL and SVM & \cite{arthur2019detecting}\\
\cline{2-4}

& Spoofing & SVM & \cite{panice2017svm}\\

\hline
\end{tabular}
}
\label{papers}
\end{table*}
Mitchell and Chen \cite{mitchell2013adaptive} analyzed attackers' behaviors according to their recklessness, randomness, and opportunistic characteristic and derived a set of behavior rules to identify attacks on the UAV. The proposed model achieved a promising detection accuracy. Moreover, it provided the capability to adjust detection strength, which allowed them to trade false-positive and false-negative rates.
Sedjelmaci \textit{et al.} \cite{sedjelmaci2017hierarchical} designed rule-based algorithms for five most lethal attacks to UAV network.
%: GPS spoofing, jamming, false information dissemination, and gray / blackhole attacks. 
They investigated how each attack impacted the network indicators, such as the signal strength intensity (SSI) or the number of packets sent (NPS).
Four rule-based detection models were implemented, and they showed a precise detection ability with low false positives in a simulated environment. Zhang \textit{et al.} \cite{zhang2018network} suggested an IDS as a hybrid model of spectral analyses. They used wavelet analysis to leverage spectral characteristics of the network traffic. They also proposed a controller and an observer tracking the traffics of the attacker to establish a precise IDS on the UAV. However, rule-based approaches were not sustainable in a different environment. Proposed methods could not sustain its detection performance when the platform or the system configuration changes. If the UAV gets updated, several rules might not be suitable for the new system. Furthermore, the detection model necessitates a more precise performance to be implemented in the real world.

To overcome the drawback, several studies applied machine learning models to detect intrusions on the UAV. Tan \textit{et al.} \cite{tan2019intrusion} applied a Deep Belief Network (DBN) with Particle Swarm Optimization (PSO). They interpreted an intrusion detection task as a massive and complicated problem. They trained a classifier with the DBN and utilized the PSO optimizer to obtain an optimal number of hidden layer nodes for the classification. DBN-PSO effectively leveraged the machine learning model and showed a significant performance rather than prior approaches. Arthur \cite{arthur2019detecting} discovered the connection of the UAV often became intermittent and left a non-linear log data. The model employed a Self-Taught Learning (STL) to gain a set of features from the flight data and utilized the Support Vector Machine (SVM) as a classifier. The proposed IDS verified its efficiency with a significant detection performance. Panice \textit{et al.} \cite{panice2017svm} utilized a SVM on the estimated state of the UAV to detect GPS spoofing attacks on UAVs. They utilized estimated states of the UAV as key features and classified the UAV status into two cases: safe status and unsafe. They achieved a promising intrusion detection performance through the binary classification of safe status and unsafe status.

The machine learning approaches demonstrated their efficacy in many studies, but they accompanied the limit of supervised learning. As proposed machine learning approaches employed supervised learning models, the practitioner must provide a well-labeled data at the training phase. In the context of the IDS on the UAV, labels indicate whether the flight data is benign or under attack, and the type of attack techniques. However, labeling every flight data requires an enormous effort and the cost. Furthermore, an IDS with supervised learning cannot recognize attacks that were not trained before. As the supervised learning models can only identify learned attacks, the IDS might be neutralized with an unseen attack into the system.

Considering analyzed drawbacks of both rule-based and machine-learning-based approaches, we propose an IDS on the UAV with unsupervised learning. As unsupervised learning does not necessitate solid labels during the model training, it reduces the burden of labeling cost to the practitioner. Furthermore, unsupervised learning enables the model to detect various intrusions that are not labeled or not pre-known. The following sections further provide a detailed description of how we designed an IDS on the UAV leveraging the efficacy of unsupervised learning.

%%%%%%%%%%%%%%%%% PROPOSED METHODOLOGY
\section{Proposed Methodology}
\subsection{Dataset}
\subsubsection{Description}
We utilized Hardware-In-The-Loop (HITL) UAV DOS \& GPS Spoofing Attacks on MAVLINK dataset \cite{uav_3} on the experiment. The dataset contains system logs along with the simulated flight. These system logs are collected under the simulated environment, which follows standard jMAVSim setup. The dataset contains system logs at the UAV under three circumstances described below:

\begin{itemize}
    \item \textbf{Benign Flight}: A log data during the flight without any attacks on the system
    \item \textbf{DoS Attack}: A log data during the flight with Denial-of-Service (DoS) attack for 11 seconds
    \item \textbf{GPS Spoofing Attack}: A log data during the flight with GPS Spoofing attack for 28 seconds
\end{itemize}

Our key takeaway of the study is utilizing the benign flight data only at the model training stage, and whether the trained model can recognize the intrusion unless attacks are not trained before. As the dataset includes both the benign flight and the flight under attacks, we analyzed we can utilize the dataset to train the model with the benign flight and validate the model with logs under two attacks: DoS attack and GPS Spoofing attack.

%\vspace{-0.3cm}
\subsubsection{Ground-truth confirmation}
We confirmed the ground-truth of the dataset by checking the timestamp of the log data. As HITL DOS \& GPS Spoofing Attacks dataset provides a particular timestamp of attack, we labeled the log between attack start time and the attack end time as the flight under attack. The log from the benign flight does not contain both attack start time and the attack end time as the flight does not include any intrusions to the system. Detailed timestamps are described in Table \ref{dataset_time}.

\begin{table*}[ht]
\centering
\caption{Particular timestamps of the dataset}\label{tab1}
\resizebox{0.95\textwidth}{!}{
\begin{tabular}{|c|c|c|c|c|}

\hline
\textbf{Dataset} & \textbf{Flight Start Time} & \textbf{Attack Start Time} & \textbf{Attack End Time} & \textbf{Flight End Time}\\
\hline

Benign Flight & 14:00:52 & - & - & 14:25:50\\
DoS Attack & 15:29:06 & 15:54:09 & 15:54:20 & 15:55:09\\
GPS Spoofing Attack & 15:58:19 & 16:24:14 & 16:24:42 & 16:26:25\\

\hline
\end{tabular}
}
\label{dataset_time}
\end{table*}

\subsection{Feature Extraction}
The dataset contains a wide range of features related to the UAV. These features are written in a system log to record the status of the UAV during the flight. We categorized every features of the dataset into five types as summarized in Table \ref{feature_types}.

\begin{table*}[ht]
\centering
\caption{Five categories of features in the dataset}\label{tab1}
\resizebox{0.92\textwidth}{!}{
\begin{tabular}{|c|l|}

\hline
\textbf{Category} & \textbf{Description} \\
\hline
Location & A set of features related to the location of the UAV. A particular coordinates\\

& of the location is described along with the Global Positioning System (GPS).\\

Position \& Orientation & A set of features related to the position and the orientation of the UAV.\\

Internal Measurements & A set of features extracted from the Internal Measurement Units (IMUs).\\

System Status & A set of features related to the system management such as on-board sensors.\\

Control & A set of features illustrating an input toward the actuator to move the UAV.\\

\hline
\end{tabular}
}
\label{feature_types}
\end{table*}

From the five categories of the feature, we extracted features that can effectively recognize abnormal patterns of the UAV during the flight under attack. We established two rules for feature extraction. First and foremost, we considered a hardware generality to select the category of the feature. Furthermore, we investigated sensor stability to choose particular features under the category. A detailed explanation is elaborated below.

%\vspace{-0.2cm}
\subsubsection{Hardware Generality}
We analyzed features shall exist regardless of the type of the UAV; thus, we extracted features related to the geographic properties and physical properties. One of the key takeaways of our study is a generality; that our models can be easily established regardless of the hardware. If a particular feature exists only at our employed UAV, the proposed model cannot be utilized at other types of UAVs. Therefore, we excluded every unique feature which only exists at our UAV (MAVLINK). For instance, we excluded features in a control category as the control input varies along with the hardware. As an input toward the actuator differs from the hardware configuration, we analyzed features in the control category that cannot be widely utilized. Instead, we selected features related to the geographic properties and physical properties as we intuitively inferred most UAV systems measure these properties during the flight. Therefore, we employed features in a geographic category - location - and physical category - position \& orientation, internal measurements, and system status.

%\vspace{-0.2cm}
\subsubsection{Sensor Stability}
We inferred selected features should not be frequently lost during the flight; therefore, we employed features that do not contain any failure from the sensor. The absence of a particular feature causes damage to the model. If a particular feature contains any missed values, this feature exercises a negative influence on the model training and inference. Moreover, a feature without any changes can blur the pattern of UAVs during the flight. The model should learn unique characteristics of benign flight; however, a tranquil feature without any changes would blur these characteristics. Therefore, we established two conditions described below and dropped every feature under any of the illustrated conditions.

\begin{itemize}
    \item \textbf{Missing Value: }A feature contains any missing values during the flight at both benign flights and the flight under attack (i.e., Null)
    \item \textbf{Tranquil Value: }A feature only includes the same value without any changes at both benign flights and the flight under attack
\end{itemize}

By considering the aforementioned hardware generality and the sensor stability, we extracted features from the dataset. The used features are described at the Table \ref{feature_set}.

\begin{table}[ht]
\centering
\caption{Features used in the analysis}\label{tab1}
\resizebox{0.98\textwidth}{!}{
\begin{tabular}{|c|c|l|}

\hline
\textbf{Category} & \textbf{Feature Name} & \textbf{Description} \\

\hline
Location & Latitude & A value of latitude from the virtual GPS system\\
& Longitude & A value of longitude from the virtual GPS system\\
& Altitude & A value of altitude from the virtual GPS system\\
& EPH & A hotizontal dilution of the position at the virtual GPS system\\
& EPV & A vertical dilution of the position at the virtual GPS system\\
& Velocity & A ground speed at the virtual GPS system\\
& Course Over Ground & A direction of the movement recorded in the angular degree\\

\hline
Position \& Orientation & Local Position (x,y,z) & Local position of the UAV in the local coordinate frame\\
& & along with the axis x,y,z, respectively\\

& Ground Speed X & Ground X speed toward the latitude, positive north\\
& Ground Speed Y & Ground Y speed toward the longitude, positive east\\
& Ground Speed Z & Ground Z speed toward the altitude, positive down\\

& Roll & A roll angle\\
& Pitch & A pitch angle\\
& Yaw & A yaw angle\\

& Roll Speed & An angular speed at the roll\\
& Pitch Speed & An angular speed at the pitch\\
& Yaw Speed & An angular speed at the speed\\

& Relative Altitude & An altitude above the home position\\
& Local Altitude & An altitude in the local coordinate frame\\

& Quaternion (1,2,3,4) & Quaternion component of w,x,y,z, respectively\\

\hline
IMUs & Acceleration (x,y,z) & An acceleration at axis x,y,z, respectively\\
& Angular Speed (x,y,z) & An angular speed around axis x,y,z, respectively\\
& Magnetic Field (x,y,z) & A value of magnetic field at at axis x,y,z, respectively\\
& Absolute Pressure & An absolute pressure at the UAV\\
& Pressure Altitude & A value of the altitude calculated from the pressure\\

\hline
System Status & Temperature & A temperature of the battery\\
& Air Speed & Current indicated airspeed\\
& Heading & Current heading in a compass units scaled in 0 to 360\\
& Throttle & Current setting of the throttle scaled in 0 to 100\\
& Climb Rate & Current level of the climb rate\\

\hline
\end{tabular}
}
\label{feature_set}
\end{table}

\subsection{Feature Engineering}
Although we extracted essential features from the dataset, we figured out two obstacles to provide the data into the model: Different scales of each feature and different periods of each feature. A different scale and periods of each feature cause a negative influence at training deep neural networks. We mitigated these two obstacles with the following feature engineering steps: Feature scaling and timestamp pooling.

%\vspace{-0.3cm}
\subsubsection{Feature Scaling}
We transformed the values of each feature under the same scale. As each feature has a different magnitude of the scale, a deep neural networks-based model would get confused easily when it optimizes parameters. If several features have much larger value than other features, the loss cannot be minimized along with the training steps; thus, it creates an obstacle at the model training. By applying the scaling function elaborated in Equation (\ref{eq:min max nomalization}), we scaled each feature under the same scope and mitigated different scales of each feature.

\begin{equation}
\label{eq:min max nomalization}
X_{scaled} = \frac{X_i - Min(X_i)}{Max(X_i) - Min(X_i)}
\end{equation}

\subsubsection{Timestamp Pooling}
\begin{figure}[h]
\centering
%\begin{tabular} {c c}\\
\includegraphics[width=8cm]{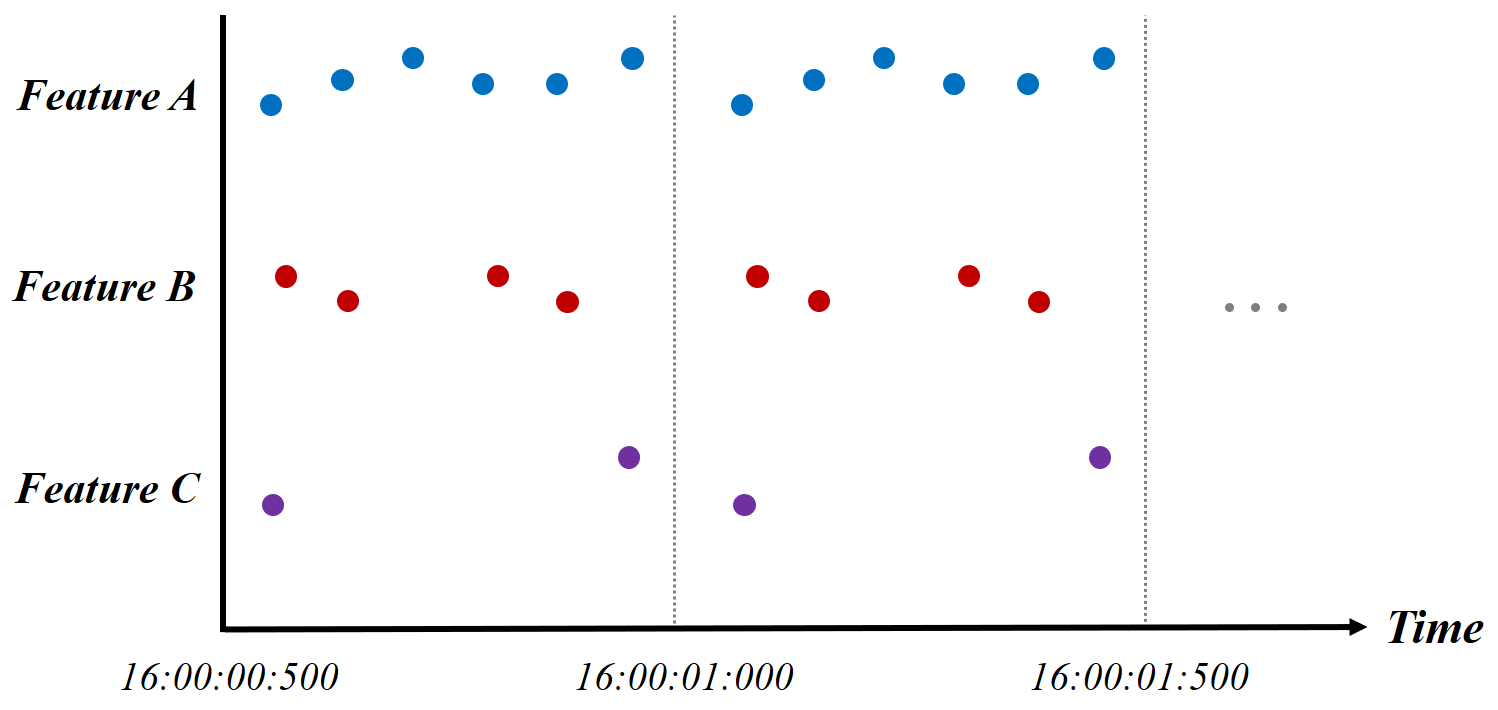} \\
(a) Before timestamp pooling \\
\includegraphics[width=8cm]{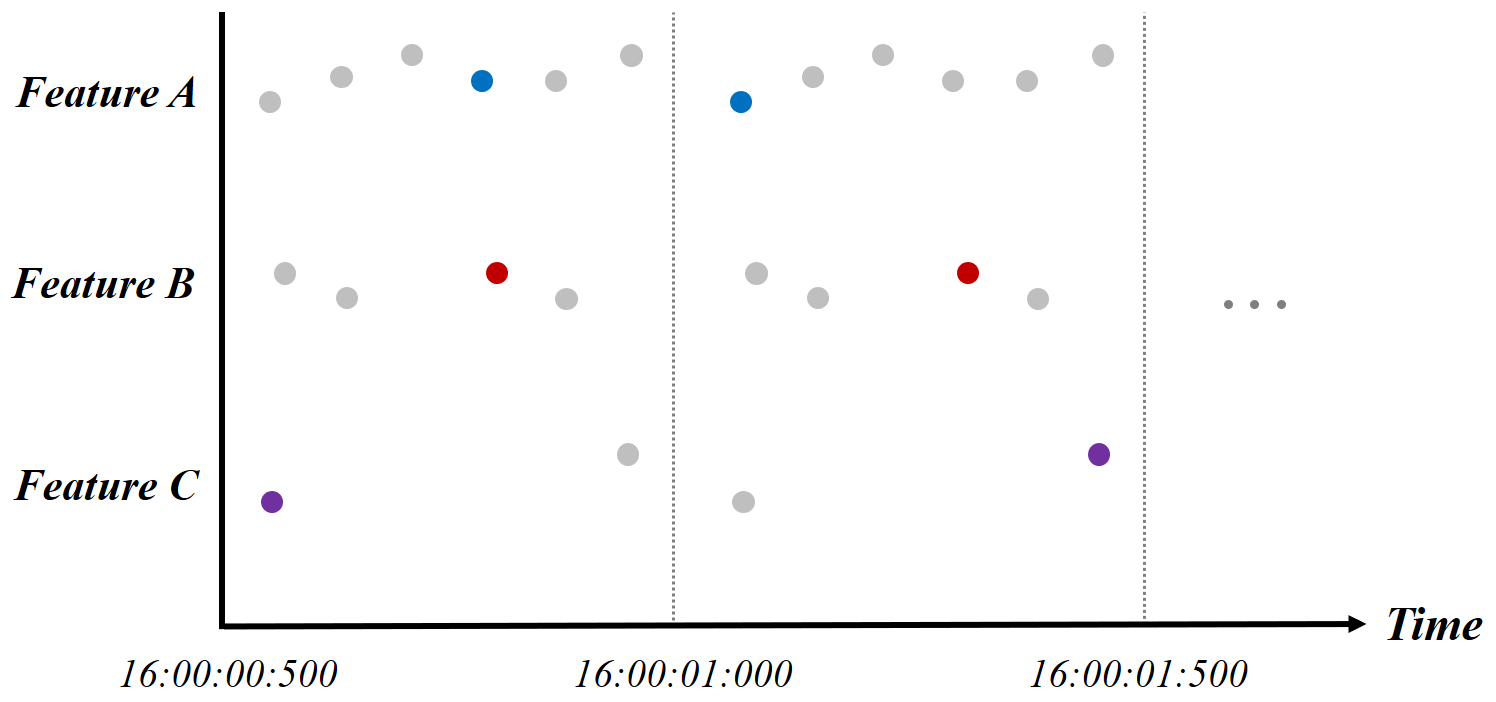} \\
(b) After timestamp pooling\\
%\end{tabular}
\caption{Timestamp Pooling}
\label{calib}
\end{figure}

We unified the length of each feature through the timestamp pooling. Following the characteristic of UAV, each feature is recorded in a different period, as visualized in Fig. \ref{calib}. (a). Referring the Fig. \ref{calib}. (a), Feature A, B, and C have different periods of data recording under the same time window. When we transform these features during a particular time window, the length of feature vectors varies. During the same time window, the number of data points at each feature is 6, 4, 2 for Feature A, B, C, respectively. As the intrusion detection system identifies the attack per timestamp, a different number of data points during the same time window become an obstacle against the model training. We interpret that each feature necessitates a transformation process with the same number of data points during the fixed time window.

To fulfill this requirement above, we applied a timestamp pooling, which is selecting a single value from the values during a fixed time window. We randomly sampled a single value from each feature, and inferred randomly-selected value can be a representative value during the fixed time window. We set the time window as 500 milliseconds and applied a timestamp pooling to every feature. If we apply the timestamp pooling at the example as mentioned earlier, the result is displayed in Fig. \ref{calib}. (b). Each feature has a single value during 500 milliseconds; thus, each feature's length has become unified. Therefore, we mitigated a different period of each feature by applying the timestamp pooling.

Throughout two feature engineering processes, we transformed raw log data into the trainable feature vector. The feature vectors have the same scope of scale and the same length; thus, we analyzed deep neural networks based model can effectively learn the pattern of the UAV.
\subsection{Unsupervised Intrusion Detection Model}
Our key intuition of the intrusion detection model is training the autoencoder to learn the dynamics of benign flights only. Then, the autoencoder trained only with benign patterns generates a low reconstruction loss toward the benign flight and high reconstruction loss with the flight under attack.

\subsubsection{Autoencoder}
The autoencoder is a deep neural network which learns the representation of feature vectors by iterating an encoding phase and a decoding phase. The encoder optimizes its parameters to find effective representations of feature vectors while the decoder optimizes its parameters to reconstruct the original input vector from the created representation \cite{schmidhuber2015deep}. A loss function of the autoencoder is defined as the difference between the original input vector and the reconstructed input vector. The autoencoder optimizes its parameters to minimize the loss; thus, a well-trained autoencoder reconstructs the input vector without much loss. In this study, we employed a linear autoencoder in which the encoder and the decoder are designed with linear neurons, and the activation function as ReLU function \cite{agarap2018deep}. For a clear elaboration of the model, we described a single layer of the encoder, single layer of the decoder, reconstruction loss (loss function), and the objective function at (\ref{encoder}), (\ref{decoder}), (\ref{loss}), and (\ref{obj}), respectively.

\begin{equation}
    Encoder: e(x)= ReLU(W_{enc}x + b_{enc})
\label{encoder}
\end{equation}

\begin{equation}
    Decoder: d(r) = ReLU(W_{dec}r + b_{dec})
\label{decoder}
\end{equation}

\begin{equation}
%\begin{split}
    Loss: L(x,y) = ||f_{\theta}(x)-y||^2\quad where,\quad f_{\theta}(x)=d^n(e^n(x))
%\end{split}
%    Loss: L(x,y) = ||f_{\theta}(x)-y||^2 where f_{\theta}(x)=d^n(e^n(x))
\label{loss}
\end{equation}

\begin{equation}
    \theta^* = argmin_{\theta}(\sum_{x\in D}L(x,y))
    \label{obj}
\end{equation}

\subsubsection{Modeling}
A key intuition of our detection models is as follows: The linear autoencoder densely trained only with benign feature vectors would produce small reconstruction loss at benign flights, but generate large reconstruction loss at abnormal flights under attack. To leverage the efficiency of unsupervised learning, we provided feature vectors from benign flights only. Note that there were no labeled feature vectors from the flight under attack at the training stage. We `densely' trained the autoencoder with the benign feature vectors only, then the parameters are optimized to reconstruct patterns from the benign flight. In other words, a well-trained autoencoder reconstructs benign feature vectors without much loss. On the other hand, the trained autoencoder will produce larger reconstruction loss with feature vectors under attack. As the autoencoder did not learn patterns of the attack, parameters are not optimized to feature vectors under attack; thus, the model produces a large reconstruction loss. Following the aforementioned intuition, we inferred the difference in reconstruction loss could be utilized to recognize the intrusion. If the data point from a particular time window records a large reconstruction loss, we can identify the existence of intrusion. Along with the experiments described in a further Section, we proved the trained autoencoder generates small reconstruction loss with benign feature vectors while it produces large reconstruction loss at feature vectors under attack.

\section{Experiment}
Our experiment's objective is to validate whether the proposed methodology effectively recognizes the intrusion from the benign flight. Throughout the experiment, we aimed to validate two key takeaway. First, we checked whether the trained model provides a larger reconstruction loss during the flight under attack rather than the benign flight. Second, we explored the difference between reconstruction losses from both the benign flight and the flight under attack. The following contents describe how we configured the experiment, and the experiment results showed the proposed model can be utilized to detect intrusion on the UAV.

%\begin{itemize}
%    \item \textbf{Takeaway 1: } Checking whether the trained model provides a larger reconstruction loss during the flight under attack
%    \item \textbf{Takeaway 2: } Explore the difference between reconstruction losses from both the benign flight and the flight under attack  
%\end{itemize}

%The following contents describe how we configured the experiment, and the experiment results showed the proposed model can be utilized to detect intrusion on the UAV.

\subsection{Setup}
We leveraged three log data (benign flight, DoS attack, and GPS Spoofing attack) from the dataset. As our approach highlights the advantage of unsupervised learning, we configured the training set only with the feature vectors from the benign flight. On the other hand, we configured two test sets from the DoS attack log data and GPS Spoofing log data. We randomly selected a particular timestamp as a starting point for the test set configuration, where the chosen timestamp is located before the attack. From this starting point, we extracted every log data until the attack ends. In this way, we configured the test set to include patterns from both benign status and the status under attack. In other words, two test sets - DoS attack and GPS Spoofing attack - have patterns from the benign status and the status under attack at the same time. After we set the training set and the test set, we applied the aforementioned feature engineering process. Note that we scaled features in the test set with the scaler used in the training set.

\subsection{Experiment Result}
\begin{figure*}[h]
\centering
%\begin{tabular} {c c}\\
\includegraphics[width=8cm]{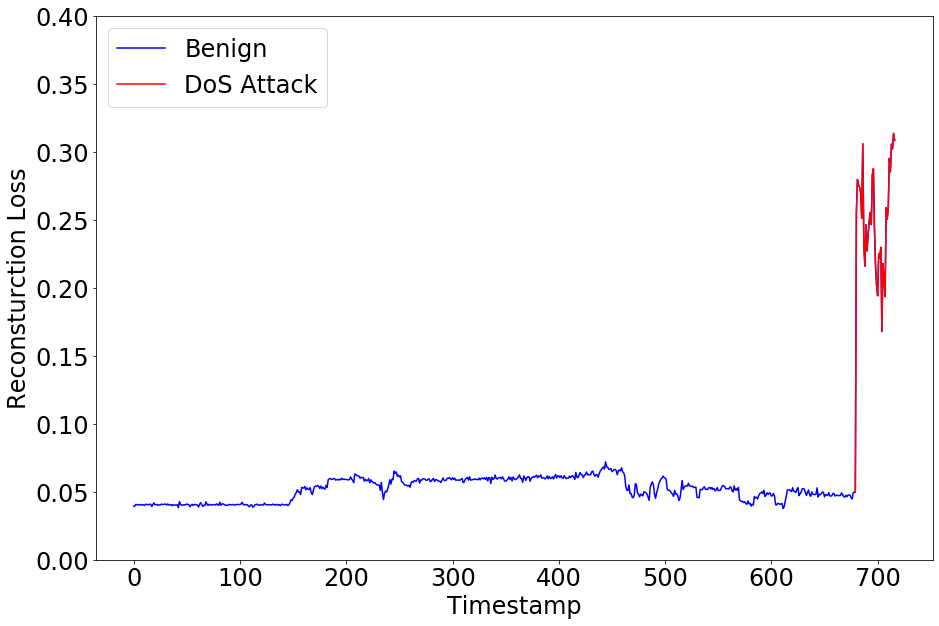} \\
(a) Flight under DoS Attack \\
\includegraphics[width=8cm]{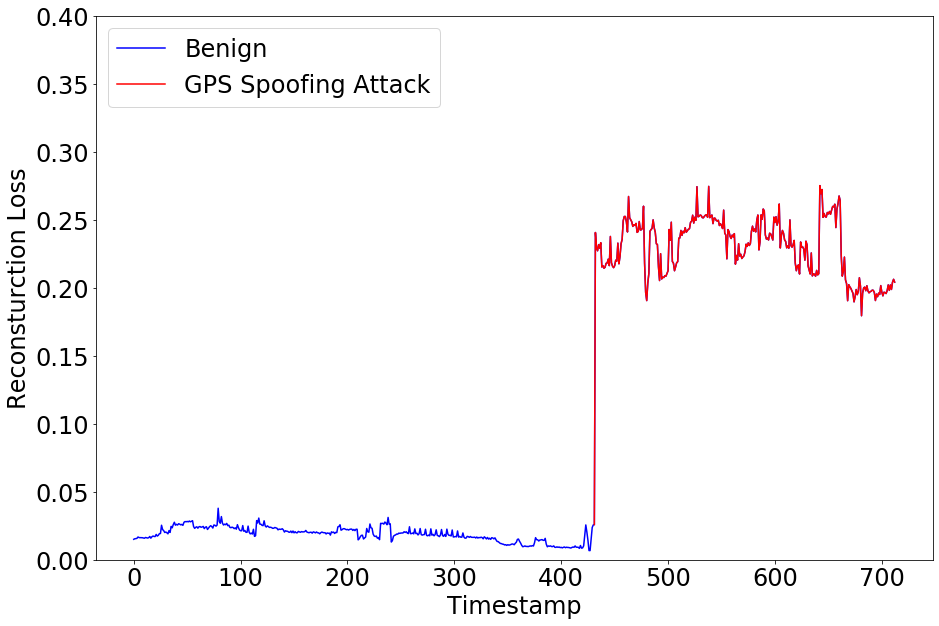} \\
(b) Flight under GPS Spoofing Attack\\

%\end{tabular}
\caption{Experiment result at two simulated flights: A flight under DoS Attack and the flight under GPS Spoofing Attack}
\label{lineplot_result}
\end{figure*}

\begin{figure}[h]
\centering
\begin{tabular} {c c}\\
\includegraphics[width=5cm]{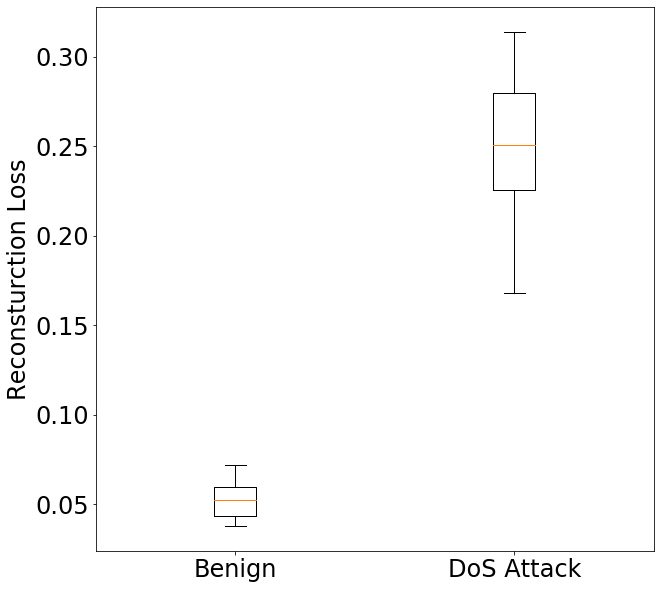} &
\includegraphics[width=5cm]{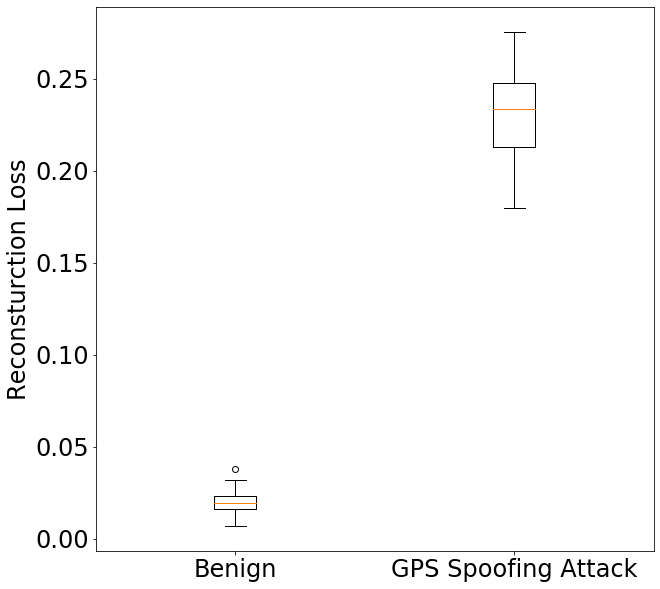} \\
(a) Flight under DoS Attack & (b) Flight under GPS Spoofing Attack\\

\end{tabular}
\caption{Reconstruction losses at both normal status and intruded status}
\label{boxplot_result}
\end{figure}

We trained a linear autoencoder with the training set, which is composed of the benign feature vectors only. To fit benign patterns into the model, we leveraged several techniques toward the model training. A batch normalization is applied toward the encoder and the decoder. We utilized both L1 regularizer \cite{zou2005regularization} and L2 regularizer \cite{cortes2012l2} to evade an overfitting problem, and parameters are optimized with Adam optimizer for an effective model training. After the model is fully trained, we provided two test sets to the model and collected reconstruction losses. The experiment results from the test of DoS attack, and the GPS Spoofing attack is described in Fig. \ref{lineplot_result} and Fig. \ref{boxplot_result}.

Fig. \ref{lineplot_result} explains the first takeaway of our experiment. The blue part of the figure implies a reconstruction loss under the benign status, and the red part of the figure stands for the loss under attack. We figured out the reconstruction losses excessively rise when the flight is under attack at both DoS attack and the GPS Spoofing attack. The reconstruction loss increases in a large amount when the feature vector from the flight under attack is provided—furthermore, Fig. \ref{boxplot_result} shows the second takeaway of the experiment is also valid. Fig. \ref{boxplot_result} illustrates a distribution of reconstruction losses at both benign status and the status under attack. The reconstruction loss distributes such far from the benign status at both DoS attack and GPS Spoofing attack. A significant difference implies a large difference in a pattern; thus, we discovered our model effectively learned the dynamics of benign patterns and recognized any abnormal patterns on the UAV. Despite a significant performance of our model, however, we figured out a room for improvement with the consideration of real-world deployment. Detailed contents are elaborated in the following section.

%%%%%%%%%%%%%%%%% DISCUSSION
\section{Discussions}
%\vspace{-0.2cm}
\subsubsection{Computation Overhead}
First, future studies can consider the computation overhead of the proposed model. Although our model produced a precise intrusion detection result, it should accompany small computing resources. Under the heavy computation overhead, the model cannot be deployed into individual UAVs as the computing environment of the UAVs is not sufficient. If the model requires substantial computing resources, it can be transformed into a lightweight. We expect future studies to reduce the computation overhead by minimizing the size of feature vectors or applying model compression techniques \cite{cheng2017survey} into the proposed model.
%\vspace{-0.2cm}

\subsubsection{Model Improvement in the Wild}
The model shall be improved with the actual flight data. As our model is trained and validated with a simulated dataset, log data would have fewer noises rather than the actual data. We expect an actual flight would be interfered with by various factors such as sensor errors, climate, and electric communication environment. The model might necessitate additional feature engineering processes to make the model learns the dynamics of benign flight. In a future study, we would collect the actual data from the UAV in the wild and improve the proposed model.
%\vspace{-0.2cm}

%%%%%%%%%%%%%%%%% CONCLUSION
\section{Conclusion}
%\vspace{-0.2cm}
An IDS is one of the key factors of the UAV safety, as it can identify an abnormal status of the system at first. Prior studies have proposed numerous approaches regarding the IDS, but they accompany limits. The rule-based models could not precisely recognize attacks during the flight. Moreover, the machine learning-based models required a great effort on data labeling, and the model could not recognize the attack which was not trained. These limits were a room for the improvement to build a practical IDS on the UAV.

We presented a novel IDS on the UAV to improve the limits of previous studies. Our study proposed an IDS leveraging an autoencoder, a deep neural network of unsupervised learning. Throughout the study, we presented a series of analyses to extract features from the raw UAV flight data. Furthermore, we trained the model only with the benign flight data and validated the model effectively recognize DoS attacks and GPS Spoofing attack though these patterns are not trained. Our model with the unsupervised learning provided two advantages. First, the model does not necessitate a heavy effort on data labeling. Second, our model can identify attacks during the flight, although the model did not learn the dynamics of the flight under attack. We expect our study can be a concrete base in the pursuit of safe utilization of UAVs in the real world.

\section{Acknowledgement}
This work was supported by Institute of Information \& communications Technology Planning \& Evaluation (IITP) grant funded by the Korea government (MSIT) (No.2018-0-00232, Cloud-based IoT Threat Autonomic Analysis and Response Technology).

%\bibliographystyle{splncs04}
%\bibliography{reference.bib}
%\input{original.bbl}

\end{document}